\documentclass[a4,a4wide]{article}
\usepackage{aaai}
\usepackage{times}
\usepackage{helvet}
\usepackage{courier}
\usepackage{color}
\usepackage{multicol}

\usepackage{amssymb}
\usepackage{graphicx}
\usepackage{color}
\usepackage{hhline}
\usepackage{pdfsync}
\newcommand{\ignore}[1]{}

\newcommand{\boxtheorem}{\hfill $\Box$\\}
\newcommand{\nit}[1]{{\it #1}}

\newcommand{\IC}{\nit{IC}}

\def\IC{{\textit{IC}}}
\usepackage{graphicx}

\usepackage{algpseudocode}
\usepackage{algorithm}

\newcommand{\comlb}[1]{{\vspace{2mm}\noindent \red{\bf COMM(LEO):}}~ #1 \hfill {\bf    END.}\\}

\newcommand{\red}[1]{\textcolor{red}{#1}}

\newcommand{\combabak}[1]{{\vspace{4mm}\noindent \bf  COMM(Babak):}~ \red {\em  #1}\hfill {\bf END.}\\}

\ignore{\newcommand{\comlb}[1]{{\vspace{4mm}\noindent \bf COMM(LEO):}~
{\em #1}\hfill {\bf END.}\\}  }

\ignore{
\newcounter{lemmaA-counter}
\newcounter{propositionA-counter}
\setcounter{lemmaA-counter}{0}

\setcounter{propositionA-counter}{0}

{\vskip \abovedisplayskip \refstepcounter{lemmaA-counter}%
\noindent {\bf Lemma A.\arabic{lemmaA-counter}.}}%

{\vskip \abovedisplayskip \refstepcounter{propositionA-counter}%
\noindent {\bf Proposition A.\arabic{propositionA-counter}.}}%
}

\newcommand{\defproof}[2]{{\noindent\bf Proof of #1:\
}#2 \boxtheorem}

\makeatother

\newcommand{\mc}[1]{\mathcal{ #1}}

\newcounter{theorem-counter}
\newcounter{corollary-counter}
\newcounter{lemma-counter}
\newcounter{definition-counter}
\newcounter{example-counter}
\newcounter{proposition-counter}
\newcounter{remark-counter}
\newcounter{definitionA-counter}
\newcounter{lemmaA-counter}
\newcounter{propositionA-counter}
\newenvironment{theorem}%
{\vskip \abovedisplayskip \refstepcounter{theorem-counter}%
\noindent {\bf Theorem \arabic{theorem-counter}.}}%
{\vskip \abovedisplayskip \refstepcounter{corollary-counter}%
\noindent {\bf Corollary \arabic{corollary-counter}.}}%
{\vskip \abovedisplayskip \refstepcounter{lemma-counter}%
\noindent {\bf Lemma \arabic{lemma-counter}.}}%

\newenvironment{definition}%
{\vskip \abovedisplayskip \refstepcounter{definition-counter}%
\noindent {\bf Definition \arabic{definition-counter}.}}%
\newenvironment{example}%
{\vskip \abovedisplayskip \refstepcounter{example-counter}%
\noindent {\bf Example \arabic{example-counter}.}}%
\newenvironment{proposition}%
{\vskip \abovedisplayskip \refstepcounter{proposition-counter}%
\noindent {\bf Proposition \arabic{proposition-counter}.}}%
{\vskip \abovedisplayskip \refstepcounter{remark-counter}%
\noindent {\bf Remark \arabic{remark-counter}.}}%
\title{\vspace*{-1cm} {\bf Causality in Databases: The Diagnosis and  Repair Connections}}
\author{{\bf Babak Salimi} \ \ and \ \ {\bf Leopoldo Bertossi}\\
 Carleton University, \
School of Computer Science\\  Ottawa,  Canada\\
\hspace*{1.5cm}\{bsalimi, bertossi\}@scs.carleton.ca
}
\begin{document}
%\pagenumbering{arabic}
%\pagestyle{headings}
%\pagestyle{empty}
\nocopyright
\maketitle

\begin{abstract}
In this work we establish and investigate the connections between causality for query answers in databases, database repairs wrt. denial constraints, and consistency-based diagnosis. The first two are relatively new problems in databases, and the third one is an established subject in knowledge representation. We show how to obtain database repairs from causes and the other way around. The vast
body of research on database repairs can be applied to the newer problem of determining actual causes for query answers. By formulating a causality problem as a diagnosis problem,
we manage to characterize causes in terms of a system's diagnoses.
\end{abstract}

\section{Introduction}

When querying a database, a user may not always obtain the expected results, and the system could provide some explanations. They could be useful to further understand the data or  check if the
query is the intended one. Actually,
the notion of explanation for a query result was introduced in \cite{Meliou2010a}, on the basis of the deeper concept of {\em actual causation}.

Intuitively, a tuple $t$  is a {\em cause} for an answer $\bar{a}$ to a
conjunctive query $\mc{Q}$ from  a relational database instance $D$ if there is a ``contingent" set of tuples $\Gamma$,
such that, after removing $\Gamma$ from $D$, removing/inserting $t$ from/into $D$ causes $\bar{a}$ to switch from being an answer to being a non-answer.
Actual causes and contingent tuples  are restricted to be among a pre-specified set
of {\em endogenous tuples}, which are admissible, possible candidates for causes, as opposed to {\em exogenous tuples}.

Some causes may be stronger than others. In order to capture this observation, \cite{Meliou2010a} also introduces and investigates a quantitative metric, called {\em responsibility}, which reflects the relative degree of causality of a tuple
for a query result. In applications involving large
data sets, it is crucial to rank potential causes by their responsibility \cite{Meliou2010b,Meliou2010a}.

Actual causation, as used in  \cite{Meliou2010a}, can be traced back to
\cite{Halpern01,Halpern05}, which provides  a model-based account of causation on the basis of the  {\em counterfactual dependence}\ignore{\footnote{$A$ is a cause of $B$ if, had $A$ not happened (this is the counterfactual
condition, since $A$ did in fact happen) then $B$ would not have happened}}. Responsibility was also introduced in \cite{cs-AI-0312038}, to capture the {\em degree of causation}.

Apart from the explicit use of causality, research on explanations for query results has focused mainly, and rather implicitly, on provenance
\cite{BunemanKT01,BunemanT07,Cheney09,CuiWW00,Tannen10,Karvounarakis02,tannen}, and
more recently, on provenance for non-answers \cite{ChapmanJ09,HuangCDN08}.\footnote{That is, tracing back, sometimes through the interplay of database tuple annotations, the reasons for {\em not} obtaining a possibly  expected answer to a query.}
 A close connection between
causality and provenance has been established \cite{Meliou2010a}.  However, causality is a more refined notion that identifies causes
for query results on the basis of  user-defined criteria, and ranks causes according to
their responsibility \cite{Meliou2010b}. For a formalization of non-causality-based explanations for
query answers in DL ontologies, see \cite{borgida}.

Consistency-based diagnosis  \cite{Reiter87}, a form of model-based diagnosis \cite[sec. 10.3]{struss}, is an area of knowledge representation. The main task here is, given the {\em specification} of a system in some logical formalism and a usually unexpected {\em observation}  about the system, to obtain {\em explanations} for the observation, in the form of a diagnosis for the unintended behavior.

In a different direction, a database instance, $D$, that is expected to satisfy certain integrity constraints (ICs) may fail to do so. In this case, a {\em repair} of $D$ is a
database   $D'$ that does satisfy the ICs and {\em minimally departs} from $D$. Different forms of minimality can be applied and investigated. A {\em consistent answer} to a query
from $D$ and wrt.\ the ICs is a query answer that is obtained from all possible repairs, i.e. is invariant or certain under the class of repairs. These notions were introduced
in \cite{pods99} (see \cite{2011Bertossi} for a recent survey). We should mention that, although not in the framework of database repairs, consistency-based diagnosis techniques have been applied
to restoring consistency of a database wrt. a set of ICs \cite{Gertz97}

These three forms of reasoning, namely inferring causality in databases, consistency-based diagnosis, and consistent query answers (and repairs) are all {\em non-monotonic}.
For example, a (most responsible) cause for a query result may not be such anymore after the database is updated. In this work we establish natural, precise, useful, and deeper connections
between these three reasoning tasks.

We show that inferring and computing actual causes  and responsibility in a database setting become, in different forms,
consistency-based diagnosis reasoning problems and tasks.  Informally,
a causal explanation for a conjunctive query answer can be viewed as a diagnosis, where in essence  the first-order logical reconstruction of the relational database
 provides the system description  \cite{Reiter82}, and the observation is the query answer.
Furthermore, we unveil a
strong connection between computing causes and their responsibilities for conjunctive queries, on the one hand, and computing {\em repairs} in databases
\cite{2011Bertossi}
wrt. denial constraints, on the other hand. These computational
problems can be reduced to each other.

 More precisely, our results  are as follows:
 \begin{enumerate}
 \item For a boolean conjunctive query and its associated denial constraint (which is violated iff the query is true), we establish a precise connection
 between actual causes for the query (being true) and the subset-repairs of the instance wrt.\ the constraint. Namely, we obtain causes from repairs.
 \item In particular, we establish the connection between an actual cause's responsibility and cardinality repairs wrt.\ the associated constraint.
 \item  We characterize and obtain subset- and cardinality- repairs for a database under a denial constraint in terms of the causes for the associated query being true.
 \item We consider {\em a set} of denials constraints and a database that may be inconsistent wrt.\ them. We obtain the database repairs by means of an algorithm
 that takes as input the actual causes for constraint violations and their contingency sets.
 \item We establish a precise connection between consistency-based diagnosis for a boolean conjunctive query being unexpectedly true according to a
 system description, and causes for the query being true. In particular, we can compute actual causes, their contingency sets, and responsibilities from
 minimal diagnosis.
 \item Being this a report on ongoing work, we discuss several extensions and open issues that are under investigation.
 \end{enumerate}

 %\comlb{Leave this enumeration for the end.}

\section{Preliminaries}\label{sec:prel}

We will consider relational database schemas of the form $\mathcal{S} = (U,\mc{P})$,  where $U$ is the possibly infinite
database domain and $\mc{P}$ is a finite set of database predicates of fixed arities. A database instance $D$
compatible with $\mathcal{S}$ can be seen as a finite set of ground atomic formulas (in databases aka. atoms or tuples), of the form $P(c_1, ..., c_n)$, where $P \in \mc{P}$ has arity $n$, and $c_1, \ldots , c_n \in U$.
A conjunctive query  is a formula $\mc{Q}(\bar{x})$ of the first-order (FO) logic language, $\mc{L}(\mc{S})$, associated to $\mc{S}$ of the form \ $\exists \bar{y}(P_1(\bar{t}_1) \wedge \cdots \wedge P_m(\bar{t}_m))$,
where the $P_i(\bar{t}_i)$ are atomic formulas, i.e. $P_i \in \mc{P}$, and the $\bar{t}_i$ are sequences of terms, i.e. variables or constants of $U$. The $\bar{x}$ in  $\mc{Q}(\bar{x})$ shows
all the free variables in the formula, i.e. those not appearing in $\bar{y}$. The query is boolean, if $\bar{x}$ is empty, i.e. the query is a sentence, in which case, it is true or false
in a database, denoted by $D \models \mc{Q}$ and $D \not\models \mc{Q}$, respectively. A sequence $\bar{c}$ of constants is an answer to an open query $\mc{Q}(\bar{x})$ if $D \models \mc{Q}[\bar{c}]$, i.e.
the query becomes true in $D$ when the variables are replaced by the corresponding constants in $\bar{c}$.

An integrity constraint is a sentence of language $\mc{L}(\mc{S})$, and then, may be true or false
in an instance for schema $\mc{S}$. Given a set $\IC$ of ICs, a database instance $D$ is {\em consistent} if $D \models \IC$; otherwise it is said to be {\em inconsistent}.
In this work we assume that sets of ICs are always finite and logically consistent.
A particular class of  integrity constraints (ICs) is formed by {\em denial constraints} (DCs), which are sentences  $\kappa$ of the from:
 $\forall \bar{x} \neg  (A_1(\bar{x}_1)  \wedge \cdots  \wedge A_n(\bar{x}_n)$, where $\bar{x}= \bigcup \bar{x}_i$ and each $A_i(\bar{x}_i)$ is a database atom, i.e. predicate $A \in \mc{P}$.
DCs will receive special attention in this work. They are common and natural in database applications since they disallow combinations of database atoms.

\ignore{
\comlb{Are we going to consider built-ins in this paper?}
\combabak{ No we don't. We frequently use denial constraints therefore, I thought you might want to define them in this section so I added the lines in red. }
}

\paragraph{Causality and Responsibility.} \  Assume that the database instance is split in two, i.e. $D=D^n \cup D^x$,  where $D^n$ and $D^x$ denote the sets of {\em endogenous} and {\em exogenous} tuples,
 respectively.
A tuple $t \in D^n$ is called a
{\em counterfactual cause} for  a boolean conjunctive $\mc{Q}$ ,  if $D\models \mc{Q}$ and $D\smallsetminus \{t\}  \not \models \mc{Q}$.  A tuple $t \in D^n$ is an {\em actual cause} for  $\mc{Q}$
if there  exists $\Gamma \subseteq D^n$, called a {\em contingency set},  such that $t$ is a counterfactual cause for $\mc{Q}$ in $D\smallsetminus \Gamma$ \ \cite{Meliou2010a}.

 The {\em responsibility} of an actual cause $t$ for $\mc{Q}$ , denoted by $\rho(t)$,  is the numerical value $\frac{1}{(|\Gamma| + 1)}$, where $|\Gamma|$ is the
size of the smallest contingency set for $t$. We can extend responsibility to all the other tuples in $D^n$ by setting their value to $0$. Those tuples are not actual causes for $\mc{Q}$.

\ignore{\comlb{Extend to all tuples in $D$ or $D^n$ only?}\\
\combabak{ $D^x$ has no causal relationship with query answer due to the definition. I guess its much natural to  extend it just for $D^n$ . However there will be no problem extending
 it to all tuple. There is not point doing that though}
 }

In \cite{Meliou2010a}, causality for non-query answers is defined on basis of sets of {\em potentially missing tuples} that account for the missing answer. Computing actual causes and their responsibilities for
non-answers becomes a rather simple variation of causes for answers. In this work we focus on causality for query answers.

\begin{example}\label{ex:cfex1}
Consider a database $D$ with  relations $R$ and $S$ as below, and the query $\mc{Q} :\exists x \exists y ( S(x) \land R(x, y) \land S(y))$. $D \models \mc{Q}$ and we want to find causes for $\mc{Q}$ being true in $D$ under the assumption that all tuples are endogenous.

\begin{center} \begin{tabular}{l|c|c|} \hline
$R$  & X & Y \\\hline
 & $a_4$ & $a_3$\\
& $a_2$ & $a_1$\\
& $a_3$ & $a_3$\\
 \hhline{~--}
\end{tabular} \hspace*{1cm}\begin{tabular}{l|c|c|}\hline
$S$  & X  \\\hline
 & $a_4$ \\
& $a_2$ \\
& $a_3$ \\ \hhline{~-}
\end{tabular}
\end{center}

 Tuple $S(a_3)$ is a counterfactual cause for $\mc{Q}$. If $S(a_3)$ is removed from $D$, we reach a state
 where $\mc{Q}$ is no longer an answer. Therefore, the responsibility of $S(a_3)$ is 1. Besides, $R(a_4,a_3)$ is an actual cause for $\mc{Q}$ with contingency set
$\{ R(a_3,a_3)\}$. If $R(a_3,a_3)$ is removed from $D$, we reach  a state where $\mc{Q}$ is still an answer, but further removing $R(a_4,a_3)$ makes $\mc{Q}$ a non-answer. The responsibility of $R(a_4,a_3)$ is $\frac{1}{2}$, because its smallest contingency sets have size $1$. Likewise,  $R(a_3,a_3)$ and $S(a_4)$ are actual causes for $\mc{Q}$ with responsibility  $\frac{1}{2}$.
\boxtheorem
\end{example}
Now we can show that counterfactual causality for query answers is a non-monotonic notion.

\begin{example}\label{ex:cfex2} (ex. \ref{ex:cfex1} cont.)
Consider the same query $\mc{Q}$, but now the database instance
$D=\{S(a_3),S(a_4),R(a_4,a_3) \}$, with the
partition $D^n=\{S(a_4),S(a_3)\}$ and $D^x=\{ R(a_4,a_3)\}$. Both $S(a_3)$
and $S(a_4)$ are counterfactual
causes for $\mc{Q}$.

Now assume $R(a_3,a_3)$ is added to $D$ as an exogenous tuple,
i.e. $(D^x)^\prime=\{ R(a_4,a_3),
$ $ R(a_3,a_3)\}$. Then, $S(a_4)$ is no longer a counterfactual
cause for $\mc{Q}$ in $D' = D^n \cup (D^x)^\prime$: If $S(a_4)$ is removed from the database,
$\mc{Q}$ is still true in $D'$.
Moreover, $S(a_4)$ not an actual cause anymore, because there is no contingency set
 that makes $S(a_4)$ a counterfactual cause.

\ignore{
 \comlb{Was it an actual cause before the update?}
 \combabak{In fact every counterfactual cause is either an actual cause with empty contingency set.  So it was actual cause
 before the update. }
 }

Notice that, if $R(a_3,a_3)$ is instead inserted as an endogenous tuple, i.e.
$(D^n)^\prime=\{ S(a_4),S(a_3),R(a_3,a_3)\}$,  then, $S(a_4)$ is still an
actual cause for $\mc{Q}$, with contingency set $\{R(a_3,a_3)\}$.
\boxtheorem
\end{example}
The following proposition shows that the notion of actual causation is non-monotone in general.

\vspace{1mm}
\noindent {\bf Notation:} \ $\mc{CS}(D^n,D^x,\mc{Q})$ denotes the set of actual causes for BCQ  $\mc{Q}$ (being true) from instance $D=D^n \cup D^x$. When
$D^n = D$ and $D^x = \emptyset$, we sometimes simply write: $\mc{CS}(D,\mc{Q})$.

\begin{proposition}\label{pro:nonmon}
\ignore{Let $\mc{CS}(D^n,D^x,\mc{Q})$ denote the set of actual causes for a BCQ  $\mc{Q}$ (being true) from instance $D=D^n \cup D^x$.} Let $(D^n)^\prime, (D^x)^\prime$ denote updates of instances $D^n,D^x$ by insertion of tuple $t$, resp.
It holds: \ (a) \ignore{If $t$ inserted as an endogenous tuple i.e., $(D^n)^\prime=D^n \cup \{t\}$ then,}  $\mc{CS}(D^n,D^x,$ $\mc{Q}) \ \subseteq \ \mc{CS}((D^n)^\prime,D^x,\mc{Q})$. \
(b) \ignore{If $t$ inserted as an exogenous  tuple i.e., $(D^x)^\prime=D^x \cup \{t\}$ then, } $\mc{CS}(D^n, (D^x)^\prime, $ $\mc{Q}) \ \subseteq \ \mc{CS}(D^n,D^x,\mc{Q})$  .
\boxtheorem
\end{proposition}
\ignore{
\comlb{Can we say something about how the causes' responsibilities change?}
\combabak{ Not a general statement}}
\ignore{
\begin{proof}
\noindent (a) \ Assume an arbitrary element $t' \in  \mc{CS}(D^n,D^x,\mc{Q})$. Since $t'$ is an actual cause for $\mc{Q}$ then there exist a set $\Gamma \subseteq D^n$
such that: $D^x \cup D^n \smallsetminus  \Gamma \models \mc{Q}$ and  $D^x \cup D^n \smallsetminus  (\Gamma \cup \{t\}) \not \models \mc{Q}$. Now consider $U(D)= (D^n)^\prime \cup D^x$.
Let $\Gamma'=\Gamma \cup \{t\}$ then, $D^x \cup (D^n)^\prime \smallsetminus  \Gamma' \models \mc{Q}$ and
$D^x \cup (D^n)^\prime \smallsetminus  (\Gamma \cup \{t\}) \not \models \mc{Q}$ (because, $(D^n)^\prime \smallsetminus  \Gamma' =D^n \smallsetminus  \Gamma$). So, we obtain that $t'$ is counterfactual cause for $\mc{Q}$. Consequently, $\mc{CS}(D^n,D^x,\mc{Q})$ $\subseteq \mc{CS}((D^n)^\prime,D^x,\mc{Q})$.

\noindent (b) \
 We show that inserting an exogenous tuples do not add any new actual cause for a query answer. Assume the contradiction i.e., there exists a $t \in D^n$ such that $t \in \mc{CS}(D^n, (D^x)^\prime, \mc{Q})$ but
$t \not \in \mc{CS}(D^n, D^x, \mc{Q})$. The former statement implies that: there exists a set $\Gamma \subseteq D^n$ such that $ D^n \cup (D^x)^\prime \smallsetminus \Gamma \models \mc{Q}$
and  $ D^n \cup (D^x)^\prime \smallsetminus (\Gamma \cup \{ t \}) \not \models \mc{Q}$. The latter, implies that for all $s \subseteq D^n$ (including $\Gamma$ and $\emptyset$) we have
$(a): D^n \cup D^x \smallsetminus s \not \models \mc{Q}$ or $(b):D^n \cup D^x \smallsetminus (s \cup \{ t \}) \models  \mc{Q}$. We show that both (a) and (b) lead to contradiction.
When $s=\emptyset$, (a) is equal to $D^n \cup D^x \not \models \mc{Q}$ i.e., $ D \not \models \mc{Q}$ (contradiction!). When $s=\Gamma$, (b)
is equal to $D^n \cup D^x \smallsetminus (\Gamma \cup \{ t \}) \models  \mc{Q}$ which contradict the fact that
 $ D^n \cup (D^x)^\prime \smallsetminus (\Gamma \cup \{ t \}) \not \models \mc{Q}$, since $\mc{Q}$ is a
 monotone query. \red{Therefore, $ \mc{CS}(D^n, (D^x)^\prime, \mc{Q}) \subseteq \mc{CS}(D^n,D^x,\mc{Q}) $}. \boxtheorem
\end{proof} }
Example \ref{ex:cfex2} shows that the inclusion in (b) may be strict. It is easy to show that it can also be strict for (a). This result tells us that, for a fixed query, inserting an endogenous tuples  may extend the set of actual cases, but it may shrink by inserting an endogenous tuple. It is also easy to verify that most responsible causes may not be such
anymore after the insertion of endogenous tuples.

%The example shows that inserting tuples exogenous tuples into a database may invalidate previous \red{causes (???)} for
%a query answer. However,  inserting tuples among endogenous tuples does not change \red{causes (???)} for a particular query answer.

\ignore{\comlb{What are the general claims around (non)monotonicity in relation to counterfactual and actual
causes, and endogenous vs. exogenous? Insert a proposition with precise claims, and a proof (that we may put later
in an appendix).}  }

\vspace{-3mm}
 \paragraph{Database Repairs.} \
Given a set $\IC$ of ICs,  a {\em subset-repair} (simply, S-repair) of a possibly inconsistent instance $D$ for schema $\mc{S}$  is an instance $D'$ for
$\mc{S}$ that satisfies $\IC$ and makes $\Delta(D,D')=(D \smallsetminus D') \cup( D' \smallsetminus D)$ minimal under set inclusion.
$\nit{Srep}(D,\IC)$ denotes the set of S-repairs of $D$ wrt.\ $\IC$ \cite{pods99}. $\bar{c}$ is a {\em consistent answer} to query $\mc{Q}(\bar{x})$ if $D' \models \mc{Q}[\bar{c}]$ for every $D' \in \nit{Srep}$,
denoted $D \models_S \mc{Q}[\bar{c}]$. S-repairs and consistent query answers for DCs were investigated in detail \cite{Chomicki05}. (Cf. \cite{2011Bertossi} for more references.)

Similarly, $D'$ is a  {\em cardinality repair} (simply C-repair) of $D$ if $D'$ satisfies $\IC$ and minimizes $|\Delta(D,D')|$.
$\nit{Crep}(D,\IC)$ denotes the class of C-repairs of $D$ wrt.\ $\IC$. That $\bar{c}$ is a consistent answer to $\mc{Q}(\bar{x})$ wrt.\ C-repairs is denoted by $D \models_C \mc{Q}[\bar{c}]$.
C-repairs were investigated in detail in \cite{icdt07}.

C-repairs are S-repairs of minimum cardinality, and, for DCs,  they are obtained from the original instance by deleting a cardinality-minimum or a subset-minimal set of tuples, respectively.
Obtaining repairs and consistent answers is a non-monotonic process. That is, after an update of $D$ to $u(D)$, obtained by tuple insertions, a repair or a consistent answer for $D$ may not be such for $u(D)$ \cite{2011Bertossi}.

\ignore{

\noindent Among various types of constraints, the class of {\em denial constraints} will be used in this work. A denial constraint $\psi$  is a sentence in $L(\mathcal{S})$ of the form:
$$ \forall x \neg (A_1(\bar{x}_1) \land \ldots \land A_n(\bar{x}_n) )$$

 \noindent with $\bar{x}= \bigcup \bar{x}_i$,  $\bar{x}' \subseteq \bar{x}$ and $A_i(\bar{x}_i)$ is a database atom (i.e.,  with predicate in $P$).

\noindent Denial constraint violations can be resolved only by tuple deletions \cite{2011Bertossi}. Therefore, in this case $\Delta(D,D')$ simply becomes $(D \smallsetminus D')$.

}
%\noindent The complexity of the repair checking problem for different class of constraints and repair semantics is studied in \cite{Afrati09}. For the class of denial constraints, S-repair checking is shown in LOGSPACE while,
% C-repair checking is shown to be {\em coNP}-complete \cite{Afrati09}.

\vspace{-3mm}
\paragraph{Consistency-Based Diagnosis.} \
The starting point of this consistency-based approach to diagnosis is a diagnosis problem of the form $\mc{M}=(\nit{SD}, \nit{COMPS}, $ $ \nit{OBS})$, where $\nit{SD}$ is
the description in logic of the intended properties of a system under the {\em explicit} assumption that all its components, those in the set of constants \nit{COMPS},  are normal (or working normally).
$\nit{OBS}$ is a finite set of FO sentences (usually a conjunction of ground literals) that represents the  observations.

Now, if the system does not behave as expected (as shown by the observations),
then the logical theory obtained from $\nit{SD} \cup \nit{OBS}$ plus the explicit assumption, say $\bigwedge_{c \in \nit{COMPS}} \neg \nit{ab}(c)$, that the components are indeed behaving normally, becomes inconsistent.\footnote{Here, and as usual, the atom $\nit{ab}(c)$ expresses that component $c$ is (behaving) abnormal(ly).}
This inconsistency is captured via the {\em minimal conflict sets}, i.e. those minimal subsets $\nit{COMPS}_0$ of \nit{COMPS}, such that $\nit{SD} \cup \nit{OBS} \cup \{\bigwedge_{c \in \nit{COMPS}_0}
\neg \nit{ab}(c)\}$ is still inconsistent. As expected, different notions of minimality can be used at this point. It is common to use the distinguished predicate $\nit{ab}(\cdot)$ for denoting {\em  abnormal}
(or abnormality). So, $\nit{ab}(c)$ says that component $c$ is abnormal.

On this basis, a {\em  minimal diagnosis} for $\mc{M}$ is a minimal subset $\Delta$ of $\nit{COMPS}$, such that $ \nit{SD} \cup \nit{OBS} \cup \{\neg \nit{ab}(c)~|~c \in  \nit{COMPS} \smallsetminus \Delta \} \cup \{\nit{ab}(c)~|~c \in \Delta\}$ is consistent. That is, consistency is restored by flipping the normality assumption to abnormality for a minimal set of components, and  those are the ones considered to be (jointly) faulty. The notion of minimality commonly used is subset-minimality, i.e. a minimal diagnosis must not have a proper subset that is still a diagnosis. We will use this kind of minimality in relation to diagnosis. Diagnosis can be obtained from conflict sets \cite{Reiter87}. See also \cite[sec. 10.4]{struss} for a broader
review of model-based diagnosis.

Diagnostic  reasoning  is  non-monotonic  in  the sense  that  a  diagnosis may not survive after the addition of  new
observations \cite{Reiter87}.

\section{Repairs and Causality for Query Answers} \label{sec:caus&repair}

Let $D = D^n\cup D^x$ be a database instance for schema $\mathcal{S}$, and
$\mc{Q}\!: \exists \bar{x}(P_1(\bar{x}_1) \wedge \cdots \wedge P_m(\bar{x}_m))$  be a boolean conjunctive query (BCQ).
Suppose $\mc{Q}$ is unexpectedly true in  $D$. Actually,
it is  expected that $D \not \models \mc{Q}$, or equivalently, that $D \models \neg \mc{Q}$.
Now, $\neg \mc{Q}$ is logically equivalent to a formula of the form
$\kappa(\mc{Q})\!: \forall \bar{x} \neg (P_1(\bar{x}_1) \wedge \cdots \wedge P_m(\bar{x}_m))$, which has the form of a denial constraint.
 The requirement
that $\neg \mc{Q}$ holds can be captured by imposing the corresponding DC $\kappa(\mc{Q})$ to $D$.

Since $D \models \mc{Q}$, $D$ is inconsistent wrt.\ the DC $\kappa(\mc{Q})$. Now, repairs for (violations of) DCs are obtained
by tuple deletions.
Intuitively,  tuples that account for violations of $\kappa(\mc{Q})$ in $D$ are actual causes for $\mc{Q}$. Minimal sets of tuples like this
are expected to correspond to S-repairs for $D$ and $\kappa(\mc{Q})$. Next we make all this precise.

\ignore{
\comlb{It is important to clarify the kind of causality we are talking about. It is also, in principle confusing, that we have S-repairs on one side, a
and a notion of actual causality and responsibility that is based on cardinality-minimal contingency sets. Will this be clear? BTW, I changed in the previous version S.repairs
to S-repairs, and I asked you to change things everywhere. Do it.}
\combabak{ I will use the term ''cause" precisely from this now on.  regarding the second issue: notice that  the notion of actual causation
has noting to do with minimality of contingency-set. That is, it is good enough to show that a contingency-set exists. Actually, there is no minimality criteria for contingency sets in the definition. But for responsibility, one has to find the contingency with minimum cardinality. In the S-repair approach to causality we happen to find S-minimal contingency-sets to establish  actual cassation.}
}

 Given an instance $D = D^n \cup D^x$, a BCQ $\mc{Q}$, and a tuple $t \in D$, we consider the class containing the sets of differences between $D$ and those S-repairs that do not contain tuple $t \in D^n$, and are obtained
by removing a subset of $D^n$:
\begin{eqnarray*}
\mc{DF}(D, D^n,\kappa(\mc{Q}), t)\!\!&=&\!\!\{ D \smallsetminus D'~|~ D' \in \nit{Srep}(D,\kappa(\mc{Q})),\\&&~~~~~~~~~~~~~~~ t \in (D\smallsetminus D') \subseteq D^n\}.
\end{eqnarray*}
Now, $s \in \mc{DF}(D, D^n,\kappa(\mc{Q}), t)$ can written as  $s=s' \cup \{t\}$.
From the definition of a S-repair, including its S-minimality,   $D \smallsetminus (s' \cup \{t\}) \models \kappa(\mc{Q})$, but $D \smallsetminus s' \models \neg \kappa(\mc{Q})$, i.e.
$D \smallsetminus (s' \cup \{t\}) \not \models \mc{Q}$, but $D \smallsetminus s' \models \mc{Q}$. So, we obtain that
$t$ is an actual cause for $\mc{Q}$ with contingency set $s'$. The following proposition formalizes this result.

\begin{proposition}\label{pro:c&r}
Given an instance $D= D^n \cup D^x$, and a BCQ  $\mc{Q}$, $t \in D^n$ is an actual cause for $\mc{Q}$ iff
$\mc{DF}(D, D^n,\kappa(\mc{Q}), t) \not = \emptyset $.\boxtheorem
\end{proposition}

\noindent The next proposition shows that the responsibility of a tuple  can also be determined from $\mc{DF}(D, D^n,\kappa(\mc{Q}), t)$.

\begin{proposition}\label{pro:r&r}
Given an instance $D= D^n \cup D^x$, a BCQ  $\mc{Q}$, and $t \in D^n$,
\begin{enumerate}
\item If $\mc{DF}(D, D^n, \kappa(\mc{Q}),  t) = \emptyset$, then $\rho(t)=0$.

\item Otherwise, $\rho(t)=\frac{1}{|s|}$, where $s \in \mc{DF}(D,D^n, \kappa(\mc{Q}), t)$
and there is no $s' \in \mc{DF}(D, D^n,\kappa(\mc{Q}), t)$ such that, $|s'| < |s|$. \boxtheorem
\end{enumerate}
\end{proposition}

\ignore{
\comlb{There should be a connection with C-repairs, right?}
\combabak{ Well, I would say that the differences between $D$ and its C-repairs wrt.\ $\kappa(\mc{Q})$ are most responsible actual causes} }

\begin{example}\label{ex:CausASrepex1} (ex. \ref{ex:cfex1} cont.) Consider the same instance $D$ and query $\mc{Q}$. In this case, the DC
$\kappa(\mc{Q})$ is, in Datalog notation as a  negative rule:  \ $\leftarrow S(x),R(x, y),S(y)$.

Here, $\nit{Srep}(D, \kappa(\mc{Q}))$ $=$ $\{D_1,$ $D_2,$ $D_3\}$ and $\nit{Crep}(D, \kappa(\mc{Q}))=\{D_1\}$, with
$D_1= \{R(a_4,a_3),$ $ R(a_2,a_1),$ $ R(a_3,a_3),$ $  S(a_4),$ $ S(a_2)\}$, \ $D_2 = \{ R(a_2,a_1),$ $ S(a_4),$
$S(a_2),$ $S(a_3)\}$, \ $D_3 = \{R(a_4,a_3),$ $ R(a_2,a_1),$ $ S(a_2),$ $ S(a_3)\}$.

\ignore{
\vspace{2mm}
\hspace*{1cm} $D_1$: \begin{tabular}{l|c|c|} \hline
$R$  & X & Y \\\hline
& $a_4$ & $a_3$\\
& $a_2$ & $a_1$\\
& $a_3$ & $a_3$\\
 \hhline{~--}
\end{tabular} \hspace*{1cm}\begin{tabular}{l|c|c|}\hline
$S$  & X  \\\hline
 & $a_4$ \\
& $a_2$ \\
\hhline{~-}
\end{tabular} \hspace*{1cm} $D_2$: \begin{tabular}{l|c|c|} \hline
$R$  & X & Y \\\hline
& $a_2$ & $a_1$\\
 \hhline{~--}
\end{tabular} \hspace*{1cm} \begin{tabular}{l|c|c|}\hline
$S$  & X  \\\hline
 & $a_4$ \\
& $a_2$ \\
& $a_3$ \\ \hhline{~-}
\end{tabular}

\vspace{1mm}
\hspace*{1cm} $D_3$: \begin{tabular}{l|c|c|} \hline
$R$  & X & Y \\\hline
 & $a_4$ & $a_3$\\
& $a_2$ & $a_1$\\
 \hhline{~--}
\end{tabular} \hspace*{1cm} \begin{tabular}{l|c|c|}\hline
$S$  & X  \\\hline
& $a_2$ \\
& $a_3$ \\ \hhline{~-}
\end{tabular}
\vspace{1.5mm}
}

For tuple $R(a_4,a_3)$, $\mc{DF}(D, D, \kappa(\mc{Q}), R(a_4,a_3))=\{D \smallsetminus D_2\} = \{ \{ R(a_4,a_3),$ $R(a_3 $ $,a_3)\} \}$. This, together with Propositions  \ref{pro:c&r} and \ref{pro:r&r},
confirms that $R(a_4,a_3)$ is an actual cause,  with responsibility $\frac{1}{2}$.

For tuple $S(a_3)$,  $\mc{DF}(D, D, \kappa(\mc{Q}), $ $ S(a_3)) = \{D \smallsetminus D_1\}$ $=\{ S(a_3) \}$. So, $S(a_3)$
is an actual cause  with responsibility 1. Similarly, $R(a_3,a_3)$ is an actual cause with responsibility $\frac{1}{2}$, because  $\mc{DF}(D, D,\kappa(\mc{Q}),
R(a_3,a_3)) = \{D\smallsetminus D_2, \ D \smallsetminus D_3\}$ $=\{ \{R(a_4,$ $a_3),$ $R(a_3,a_3)\},$ $ \{R(a_3,a_3), S(a_4)\} \}$.

It is easy to verify that  $\mc{DF}(D,$ $ D, \kappa(\mc{Q}),S(a_2))$ and $\mc{DF}(D, D, \kappa(\mc{Q}),R(a_2 , a_1))$ are empty, because all repairs contain those tuples.
This means that they do not participate in the violation of $\kappa(\mc{Q})$, or equivalently, they do not contribute to make $\mc{Q}$ true.
 So, $S(a_2)$
and $R(a_2 , a_1)$ are not actual causes for $\mc{Q}$, confirming the result in Example \ref{ex:cfex1}. \boxtheorem
\end{example}
\ignore{
\comlb{Here we always end up having a single DC. Any way to extend this? Having a single DC may not reduce the applicability of complexity results
for DCs?}
\combabak{ Having a single DC is not problemistic when we use repair to address causality. because we are searching for causality an responsibility for a single
conjunctive query.  But for the other direction you are right. using causality to compute repairs is going to makes problem since in one hand we have a set of denial
constraints but on the other hand we have a single conjunctive query. \red{I will address this issue in another comment in next page.}}
\comlb{I do not need more comments, but results. See the second last item in the new section.}  }
  Now, we reduce  computation of repairs for inconsistent databases wrt.\ a denial constraint to corresponding problems for causality.

 Consider the database instance $D$  for  schema $\mathcal{S}$ and a denial constraint
 $\kappa\!: \ \leftarrow A_1(\bar{x}_1),\ldots,A_n(\bar{x}_n)$, to which a boolean conjunct
 ive {\em violation view} $V^\kappa\!: \exists\bar{x}(A_1(\bar{x}_1)\wedge \cdots \wedge A_n(\bar{x}_n))$ can be associated:  $D$ violates (is inconsistent wrt.) $\kappa$
 iff $D \models V^\kappa$.

\ignore{
\vspace*{-1mm} \begin{proposition}\label{pro:vq} \em
Given a database instance $D$ and a denial constraint $\psi$,  $D$ is inconsistent wrt.\ $\psi$ iff $D \models \nit{viol}_\psi$.
\vspace*{-1mm}
\boxtheorem
\end{proposition}  }

Intuitively, actual causes for $ V^\kappa$, together with their contingency sets, account for violations of $\kappa$ by $D$. Removing those tuples from $D$ should remove the inconsistency.

Given an inconsistent instance $D$ wrt.\  $\kappa$, we collect all S-minimal contingency sets associated with the actual cause $t$ for $V^\kappa$, as follows:
\ignore{
\comlb{We should have $D^n$ below? Where else? Are you assuming that $t \notin s$? Also, notice the right notation for strict set inclusion.}
\combabak{ Bringing the notion of exo/end  partitioning in the context of repair means to split the database such that some part of the database is assumed to be suspicious for participating in inconsistencies. That is we can take advantage of this partitioning
to restrict the repairs. If you think that this makes sense I can add this up to the paper  }
\combabak{$t \notin s$ due to the definition }
\comlb{I do not see how the definition captures that. It's only implicit.}
\combabak{ For any $ s\subseteq \red{D^n}$ foe which the following holds $t$ should not be $s$. otherwise it could not satisfy the following relashin.}    }
\begin{eqnarray*}
\mc{CT}(D,D^n,V^\kappa,t) &=& \{  s\subseteq D^n~|~D\smallsetminus s \models V^\kappa,\\
&& D\smallsetminus (s \cup \{t\}) \not \models V^\kappa,   \mbox{ and }  \nonumber \\
   &&  \forall s''\subsetneqq s, \ D \smallsetminus (s'' \cup \{t\})  \models V^\kappa \}. \nonumber
\end{eqnarray*}
Notice that for  sets $s \in \mc{CT}(D,D^n,V^\kappa,t)$, $t \notin s$.   Now consider, $t \in \mc{CS}(D, \emptyset, V^\kappa)$, the set of actual causes for $V^\kappa$ when the entire database is endogenous.  From  the definition of an actual cause and the S-minimality of sets
$s \in \mc{CT}(D,D, V^\kappa, t)$,  $s''= s \cup \{t\}$ is an S-minimal set such that $D \smallsetminus s'' \not \models V^\kappa$. So,  $D \smallsetminus s''$
is an S-repair for $D$.  We obtain:

\begin{proposition}\label{pro:sr&cp}
(a) Given an instance $D$ and a DC $\kappa$, $D$ is consistent wrt.\ $\kappa$ iff
$\mc{CS}(D, \emptyset, V^\kappa) = \emptyset$. \ (b)
$D' \subseteq D$ is an S-repair for $D$ iff,  for every $t \in D \smallsetminus D'$,
$t \in \mc{CS}(D, \emptyset, V^\kappa)$ and $D \smallsetminus (D' \cup \{t\}) \in \mc{CT}(D, D,V^\kappa, t)$.
\boxtheorem
\end{proposition}
 Now we establish a connection between most responsible actual causes and C-repairs. For this, we collect the most responsible actual causes for $V^\kappa$:
\begin{eqnarray*}
\mc{MRC}(D, V^\kappa)&\!\!=\!\!& \{t \in D~|~ t \in \mc{CS}(D,\emptyset,V^\kappa),\\ && \not \exists t' \in \mc{CS}(D,\emptyset,V^\kappa) \mbox{ with } \rho(t')> \rho(t)   \}.
\end{eqnarray*}
\begin{proposition}\label{pro:cr&mrp}
For an instance $D$ and denial constraint $\kappa$, $D'$ is a C-repair for $D$ wrt.\ $\kappa$ iff  for each $t \in D \smallsetminus D'$:
$t \in  \mc{MRC}(D,V^\kappa)$ and $D \smallsetminus (D' \cup \{t\}) \in \mc{CT}(D, D,V^\kappa, t)$.
\boxtheorem
\end{proposition}

\ignore{\comlb{In the prop. you had S-repair. I guess it should be C-repair.}}

\begin{example}\label{ex:rc2cp} Consider  $D=\{P(a,b),R(b,c), R(b,b)\}$, and the denial constraint
$\kappa\!: \ \leftarrow P(x, y),R(y, z)$, which prohibits a join between $P$ and $R$.
The corresponding violation view (query) is, $V^\kappa\!: \exists xyz (P(x, y) \land R(y, z))$.
Since $D \models V^\kappa$, $D$ is inconsistent wrt.\ $\kappa$.

 Here, $\mc{CS}(D, \emptyset, V^\kappa)=\{ P(a,b),R(b,c), R(b,b) \}$, each of whose members is  associated with S-minimal contingency sets:
 \ $\mc{CT}(D, D,V^\kappa,R(b,c))=\{\{ R(b,b)\}\}$,
$\mc{CT}(D, D, V^\kappa, R(b,b))=\{ \{R(b,c)\}\}$, and $\mc{CT}(D, D,V^\kappa,P(a,b))=\{\emptyset\}$.

According to Proposition \ref{pro:sr&cp},  the instance obtained by removing each actual cause for $ V^\kappa$ together with its contingency set forms a S-repair for $D$.
Therefore,  $D_1=D \smallsetminus \{P(a,b) \}=\{ R(b,c), R(b,b)\}$ is an S-repair. Notice that the S-minimal contingency set associated to
 $P(a,b)$ is an empty set. Likewise,   $D_2=D \smallsetminus \{R(b,c), R(b,b) \}=\{P(a,b)\}$ is a S-repair. It is easy to verify that $D$ does not have any S-repair other than $D_1$ and $D_2$.

Furthermore, $\mc{MRC}(D, V^\kappa)= \{P(a,b)\}$.  So, according to Proposition \ref{pro:cr&mrp}, $D_1$ is also a C-repair for $D$.
\boxtheorem
\end{example}

%\subsection{Causality and consistent query answering}

Given an instance $D$, a DC $\kappa$ and a ground atomic query $A$, the following proposition
establishes the relationship between consistent query answers to $A$ wrt.\ the S-repair semantics and actual cases for the violation view $V^{\kappa}$.

\begin{proposition}\label{pro:cqa}
A ground atomic query $A$, is consistently true, i.e. $D \models_S A$, iff $A \in D \smallsetminus \mc{CS}(D,\emptyset, V^{\kappa})$.
\boxtheorem
\end{proposition}
% can easily be extended to conjunction of ground atomic queries. \red{ The same things goes for negations of ground atomic queries}.

%\comlb{?????}

\begin{example}\label{ex:cqa1} Consider  $D=\{P(a,b),R(b,c),R(a,d)\}$,  the DC
$\kappa\!: \ \leftarrow P(x, y),R(y, z)$, and the ground atomic query $\mc{Q}\!: \ R(a,d)$. It is easy to see that
$\mc{CS}(D,\emptyset, V^{\kappa})=\{P(a,b),R(b,c)\}$. Then, according to Proposition \ref{pro:cqa}, $R(a,d)$ is consistently true in $D$, because
 $D \smallsetminus \mc{CS}(D,\emptyset, V^{\kappa})= \{R(a,d)\}$.
\boxtheorem
\end{example}

\ignore{
\red{ Note that in order to go beyond ground atomic queries, we need to extend the definition of causation
for query results in such a way that causes are defined in terms of attributes rather than tuples.
}

\comlb{What is this claim above? It does not make any sense.}
}

\section{Causes for IC violations}\label{sec:cauViol}

We may consider a set $\Sigma$ of ICs $\psi$ that have violation views $V^\psi$ that become boolean conjunctive queries, e.g. denial constraints.
    Each of such views has the form $V^\psi \!: \ \exists \bar{x} (A_1(\bar{x}_1) \wedge \cdots \wedge A_n(\bar{x}_n))$. When the instance $D$ is inconsistent wrt.\ $\Sigma$, some of these
    views (queries) get the answer \nit{yes} (they become true), and for each of them there is a set $\mc{C}(D,D^n,V^\psi)$ whose elements are
    of the form $\langle t, \{C_1(t), \ldots, C_m(t)\}\rangle$, where $t$
    is a tuple that is an  actual cause for $V^\psi$, together with their contingency sets $C_i(t)$,
    possibly minimal in some sense. The natural question is whether we can obtain repairs of $D$ wrt.\ $\Sigma$ from the sets $\mc{C}(D,D^n,V^\psi)$.

    In the following
    we consider the case where $D^n = D$, i.e. we consider the sets $\mc{C}(D,D,V^\psi)$, simply denoted $\mc{C}(D,V^\psi)$. We recall that $\mc{CS}(D,V^{\psi})$ denotes the set of
    actual causes for $V^{\psi}$. We denote with $\mc{CT}(D,V^\psi,t)$ the set of all  subset-minimal contingency sets associated with the actual cause $t$ for $V^\psi$.

   The  (naive) Algorithm \ignore{\ref{alg:srep},} \nit{SubsetRepairs} that we describe in high-level term in the following accepts as input an instance
    $D$, a set of DCs $\Sigma$, and the sets $\mc{C}(D,V^\psi)$, each of them with elements of the form $\langle t, \{C_1(t), \ldots, C_m(t)\}\rangle$ where each $C_i(t)$ is subset-minimal. The output of the algorithm is
    $\nit{Srep}(D,\Sigma)$, the set of S-repairs for $D$.

   The idea of the algorithm is as follows. For each $V^{\psi}$, $D \smallsetminus ( \{t\} \cup C(t))$ where, $t \in \mc{CS}(D,V^{\psi})$
    and $ C(t) \in \mc{CT}(D,V^\psi,t)$, is consistent with $\psi$ since, according to the definition of an actual cause, $D \smallsetminus (\{t\} \cup C(t)) \not \models {V_\psi}$.

    Therefore, $D'=D  \smallsetminus  \bigcup_{\psi \in \Sigma} \{ \{t\} \cup C(t)$ $|$ $t \in \mc{CS}(D,V^{\psi}) \mbox{ and } C(t) \in \mc{CT}(D,V^\psi,t)\}$ is consistent with $\Sigma$.
    However, it
    may not be an S-repair, because some violation views may have common causes.

   In order to obtain
    S-repairs, the algorithm  finds common causes for the violation views, and avoids removing redundant tuples to resolve inconsistencies. In this direction, the
     algorithm forms a set collecting all the actual causes for violation views: \
    $S = \{ t \; | \; \exists \psi \in \Sigma, t \in \mc{CS}(D,V^\psi)\}$. It  also builds the collection of non-empty sets of actual causes for each violation view: \ $\mc{C} = \{ \mc{CS}(D, V^\psi) \; | \; \exists \psi \in \Sigma, \mc{CS}(D, V^\psi)  \not = \emptyset \}$.
    Clearly,  $\mc{C}$  is a collection of subsets of set $S$.

    Next, the algorithm computes the set of all subset-minimal {\em hitting sets} of the collection $C$.\footnote{A set $S' \subseteq S$ is a hitting set for
    $\mc{C}$  if, for every  $C_i \in \mc{C}$, there is a   $c \in C_i$ with $c \in S'$. A hitting set is subset-minimal if no proper subset of
    it is also a hitting set.}  Intuitively, an S-minimal hitting set of $C$ contains an S-minimal set of actual causes that
    covers (i.e. intersects) all violation views, i.e. each violation view has an actual cause in the hitting set. The algorithm collects all S-minimal hitting sets of $C$ in $\mc{H}$.

  Now, for a hitting set $ h \in \mc{H} $, for each $t \in h$, if $t$ covers $V_\psi$, the algorithm
  removes both $t$ and $C(t)$ from $D$ (where
   $ C(t) \in \mc{CT}(D,V^\psi,t)$). Since it may happen that
   a violation view is covered by more than one element in $h$, the algorithm makes sure that just one of them
   is chosen. The result is an S-repair for $D$. The algorithm repeats
   this procedure for all sets in $\mc{H}$. The result is $\nit{Srep}(D,\Sigma)$.

\begin{example}\label{ex:al1}
Consider the instance $D=\{P(a,b),R(b,c),S(c,d)\}$, and the set of DCs $\Sigma=\{ \psi_1,\psi_2\}$,
with $\psi_1\!: \ \leftarrow P(x, y),R(y, z)$, and $\psi_2\!: \ \leftarrow R(x, y),S(y, z)$.
The corresponding violation views are $V^{\psi_1}\!:  \exists x y z(P(x, y) \wedge R(y, z))$, and
$V^{\psi_2}\!:  \exists x y z(R(x, y) \wedge S(y, z))$.

Here, $\mc{C}(D,{V^{\psi_1}})=\{ \langle P(a,b), \{\emptyset\} \rangle, \langle  R(b,c), \{\emptyset\} \rangle  \}$, and \
$\mc{C}(D, {V^{\psi_2}})=$  $\{\langle R(b,c), $ $ \{\emptyset\} \rangle, \langle  S(c,d), \{\emptyset\} \rangle  \}$.

The set $S$ in the algorithm above, actual causes for $\psi_1$ or $\psi_2$, is $S=\{P(a,b),$ $R(b,c),S(c,d)\}$. The  collection $C$,  of sets of
actual causes for $\psi_1$ and $\psi_2$, is $C= \{ \{ P(a,b), $ $R(b,c)\}, \{ R(b,c), S(c,d) \}   \}$.

The subset-minimal hitting sets for the collection $C$ are: $h_1=\{ R(b,c) \} $, $h_2=\{ S(c,d), P(a,b)  \}$. Since the contingency set for each of the actual causes is empty,
$D \smallsetminus h_1$ and $D \smallsetminus h_2$ are the S-repairs for $D$.
\boxtheorem
\end{example}

    The following theorem states that algorithm \nit{SubsetRepairs} provides a sound and complete method for computing $\nit{Srep}(D,\Sigma)$.

\begin{theorem}\label{the:cicv}
Given an instance $D$, a set $\Sigma$ of DCs, and the sets $\mc{C}(D,V^{\psi})$, for $\psi \in \Sigma$, \ $\nit{SubsetRepairs}$  computes exactly $\nit{Srep(D,\Sigma)}$.
\boxtheorem
\end{theorem}
 The connection between causality and databases repair provides this opportunity to apply results and techniques developed in each context to the other one.
In particular, in our future works we will use this connection to provide some complexity results in the context of consistent query answering.

\ignore{
\comlb{As I said in my message, it is not clear that, since you have single DCs, you will be in position to apply in causality many
interesting results for repairs wrt.\ DCs. In \cite{Chomicki05} there are some algorithms for computing repairs and CQA. Could we use them here? As better/alternative
algorithms to compute causality?}

\combabak{ Since computing causality for  conjunctive queries is in PTIME I am not sure if there is any room for an improvement using \cite{Chomicki05}. But definitely we can
use \cite{Chomicki05} for computing responsibility. Because no general algorithm provided for computing responsibility }

\comlb{Any connection between consistent query answering and causality? More generally, is there any interesting notions about query answering related to causes for (possibly other)
queries? Something to think about.}

\combabak{ Well,I guess we can say something about consistent query answering wrt.\  ``ground atomic queries" and possibly their negation. We characterized all S-repairs and C-repairs
of an inconsistent state wrt.\ denial constraints in terms of causality and responsibility. Likewise we can characterize the inconsistent data in database (union of all differences between
the database and its repairs). I am not sure how to go beyond  ground atomic queries though. The problem is causality is in tuple level but CQA would be in attribute level.
Extending causality to attributes rather than tuples would be an interesting thing to think about.
}
}

\section{Diagnosis and Query Answer Causality} \label{sec:MBDtoRep}

As before, let $D = D^n\cup D^x$ be a database instance for schema $\mathcal{S}$, and
$\mc{Q}\!: \exists \bar{x}(P_1(\bar{x}_1) \wedge \cdots \wedge P_m(\bar{x}_m))$  be BCQ.
Assume that $\mc{Q}$ is, possibly  unexpectedly, true in  $D$. Also as above, the associated
DC is $\kappa(\mc{Q})\!: \forall \bar{x} \neg (P_1(\bar{x}_1) \wedge \cdots \wedge P_m(\bar{x}_m))$.
So, it holds $D \not \models \kappa(\mc{Q})$, i.e. $D$ violates the DC. This is our observation, and
we want to find causes for it, using a diagnosis-based approach. Those causes will become causes for
$\mc{Q}$ being true; and the diagnosis will uniquely determine those causes.

\ignore{
 Actually,
it is  expected that $D \not \models \mc{Q}$, or equivalently, that $D \models \neg \mc{Q}$.
Now, $\neg \mc{Q}$ is logically equivalent to a formula of the form
$\kappa(\mc{Q})\!: \forall \bar{x} \neg (P_1(\bar{x}_1) \wedge \cdots \wedge P_m(\bar{x}_m))$\\
Violations of $D$ from $\kappa(\mc{Q})$ is an unexpected observation. We follow a
diagnostic approach to find explanations for this observation. Actually, we show that diagnoses for this observation are uniquely determine actual causes for $\mc{Q}$.
}

In this direction, for each predicate $P \in \mc{P}$, we introduce predicate $\nit{ab}_P$, with the same arity as $P$. Any tuple
in its extension is said to be {\em abnormal} for $P$. Our ``system description", $\nit{SD}$, for a diagnosis problem will include, among other elements,
the original database, expressed in logical terms, and the DC being true ``under normal conditions".

%\comlb{Notice that this DC is not conjunctive anymore. It has negation. Just in case ...}

\ignore{ Consider the diagnosis setting $\mathcal{M}=(\red{\nit{SD}(D)},D^n, \kappa(\mc{Q})^{ex})$, where $\red{\nit{SD(D)}}$ becomes our system description around instance $D$.
It contains the following elements: (a) $\nit{Th}(D)$, which is
Reiter's logical reconstruction of $D$ as a FO
theory \cite{Reiter82}. \ (b) The sentences $\forall \bar{x}(\nit{ab}_P(\bar{x}) \rightarrow \mbox{\bf false})$, and $\forall \bar{x}(\nit{ab}_P(\bar{x}) \rightarrow P(\bar{x}))$.
Here, {\bf false} is a propositional atom that is always false. So, the system description say that there are no abnormal tuples. The last statement, an inclusion dependency, connects
the abnormality predicates with their corresponding database predicates. }

More precisely, we consider the following {\em diagnosis problem},  $\mathcal{M}=(\nit{SD},D^n, \mc{Q})$, associated to $\mc{Q}$. Here, $\nit{SD}$ is the FO system description
that contains the following elements:

\vspace{1mm}
\noindent  (a) $\nit{Th}(D)$, which is
Reiter's logical reconstruction of $D$ as a FO theory \cite{Reiter82}. 

\vspace{1mm}
\noindent (b) Sentence $\kappa(\mc{Q}){^{ext}}$, which is $\kappa(\mc{Q})$ rewritten as follows:
\begin{eqnarray}
\kappa(\mc{Q}){^{ext}}\!\!\!&\!\!:& \!\!\!\!\! \forall   \bar{x}\neg (P_1(\bar{x}_1)  \wedge \neg \nit{ab}_{P_1}(\bar{x}_1)  \wedge \cdots \wedge  \label{eq:ext} \\
&&~~~~~P_m(\bar{x}_m) \wedge \neg \nit{ab}_{P_m}(\bar{x}_m) ). \nonumber
\end{eqnarray}
(This formula can be refined by applying the abnormality predicate, $\nit{ab}$, 
to endogenous tuples only.) 

\vspace{1mm}
\noindent (c) \ \ignore{The sentence $\neg \kappa(\mc{Q}) \longleftrightarrow \mc{Q}$, where $\mc{Q}$ is the initial boolean query. (d)} The inclusion dependencies: \
$\forall \bar{x}(\nit{ab}_P(\bar{x}) \rightarrow P(\bar{x}))$.\\

Now, the last entry in $\mathcal{M}$, $\mc{Q}$, is the {\em observation}, which together with \nit{SD} will produce (see below) and inconsistent theory. This is
because in $\mc{M}$ we make the initial and explicit assumption that all the abnormality predicates are empty (equivalently, that  all tuples are normal), i.e. we
consider, for each predicate $P$, the sentence
\begin{equation} \label{eq:default}
\forall \bar{x}(\nit{ab}_P(\bar{x}) \rightarrow \mbox{\bf false}),
\end{equation} where, {\bf false} is a propositional atom that is always false.
Actually, the second entry in $\mc{M}$
tells us how we can restore consistency, namely by (minimally) changing the abnormality condition of tuples in $D^n$. In other words, the rules (\ref{eq:default})
are subject to qualifications: some endogenous tuples may be abnormal. Each diagnosis for the diagnosis problem shows a subset-minimal set of endogenous tuples that are abnormal.

\begin{example}\label{ex:mbdaex5} (ex. \ref{ex:cfex2} cont.) For the instance $D=\{S(a_3),$ $S(a_4),$ $R(a_4,a_3) \}$, with $D^n$ $=$ $\{S(a_4),$ $S(a_3)\}$, consider the diagnostic problem
 $\mathcal{M}=( \nit{SD},\{S(a_4),S(a_3)\},$ $ \mc{Q})$, where
$\nit{SD}$ contains the following sentences:
\begin{itemize}
\item[(a)] Predicate completion axioms:

 $\forall x y (R(x,y) \leftrightarrow x= a_4 \wedge y = a_3)$,

 $\forall x(S(x) \leftrightarrow x = a_3 \vee x = a_4)$.

\item [(b)]Unique names assumption: $a_4 \neq a_3$.

\item [(c)] $\kappa(\mc{Q})^{ext}\!: \ \forall x y \neg ( S(x) \land \neg   \nit{ab}_S(x) \land  R(x, y) \land \neg   \nit{ab}_R(x, y) \land  S(y) \land \neg   \nit{ab}_S(y))$.

%\item [(c)] $\neg \kappa(\mc{Q}) \longleftrightarrow \mc{Q}$ \ (with $\kappa(\mc{Q}$) and $\mc{Q}$ as before).
\item[(d)] $\forall x y(\nit{ab}_R(x,y) \rightarrow R(x,y))$, \ $\forall x(\nit{ab}_S(x) \rightarrow S(x))$.
\end{itemize}
The explicit assumption about the normality of all tuples is captured by:

 $\forall x y(\nit{ab}_R(x,y) \rightarrow \mbox{\bf false})$, \ $\forall x(\nit{ab}_S(x) \rightarrow \mbox{\bf false})$. \boxtheorem
\ignore{
\begin{eqnarray}
\text{Ground atomic formulas}&:& S(a_3), S(a_4), R(a_4,a_3) \\
\text{Domain closure axiom}&:& \forall x(= (x,a_3) \lor = (x,a_4))\\
\text{Domain closure axiom}&:& \neg = (a_4,a_3) \\
\text{Equality axioms}&:&  \forall x:x=x, \forall x,y:x=y \leftrightarrow y=x, \\
  &&   \forall x,y,z:(x=y) \land (y=z) \leftrightarrow x=z\\
\text{Completion axioms}&:& \forall x(S(x) \rightarrow(= (x,a_4) \lor = (x,a_3))), \\
                       && \forall x \forall y(R(x, y)\rightarrow ((= (x, a_4) \land= (y, a_3))\\  \nonumber \\
 && \kappa(\mc{Q}){^{ex}} \leftrightarrow \nit{True}(\nit{yes})\\
 &&\neg ( \nit{True}(\nit{yes}) \land \nit{True}(\nit{no}))\\ \nonumber
\end{eqnarray}
\noindent where, $\kappa(\mc{Q})^{ex}$ is as follows:
$$ \forall y,x \neg (  S(x) \land \neg   ab_S(x) \land  R(x, y) \land \neg   ab_R(x, y) \land  S(y) \land \neg   ab_S(y)) $$
Note that the formulas in lines (1)-(8) specifies $\mc{TH}(D)$, Reiter's logical reconstruction of $D$ as a FO theory.
\\The logical theory $\mc{TH}(D) \cup  \nit{True(no)} $ and the assumption that $ \forall \bar{x}(\nit{ab}_P(\bar{x}) \rightarrow \mbox{\bf false})$ is inconsistent. This is because, $\forall \bar{x}(\nit{ab}_P(\bar{x}) \rightarrow \mbox{\bf false})$ is logically equivalent to $H: \neg ab_S(a_3) \land \neg ab_S(a_4) \land \neg ab_R(a_4,a_3)$ so we have, $\mc{TH}(D) \cup H \models \kappa(\mc{Q})^{ex}$ which inconsistent with the observation  $\nit{True(no)}$( according to rules (8) and (9)).}
\end{example}
\ignore{\comlb{Give an example here, including Reiter's theory. You can continue with it later.}
\comlb{We need the $\nit{ab}_P$s to be empty. Otherwise, there would be no
inconsistency.}
\comlb{I think it would look more elegant if we had an additional rule (in the theory) of the form $\kappa(\mc{Q}) \leftrightarrow \nit{True}(\nit{yes})$, where \nit{True} is a new unary predicate
that takes the values \nit{yes} or \nit{no} (which can be specified in the theory as well). The observation should be $\nit{True(no)}$.}
\comlb{UP TO HERE. Check/revise the rest at the light of my changes above. Take a loot at the new section below. Specially the *second last* one. We may want to bring it to the body of the paper.}
}

Now, the observation is $\mc{Q}$ (is true), obtained by evaluating query $\mc{Q}$ on (theory of) $D$. In this case, $D \not \models \kappa(\mc{Q})$. Since all the abnormality predicates are assumed to
be empty, $\kappa(\mc{Q})$ is equivalent to $\kappa(\mc{Q})^\nit{ext}$, which also becomes false wrt $D$. As a consequence,  $\nit{SD} \cup \{(\ref{eq:default})\} \cup \{\mc{Q} \}$ is an inconsistent FO theory.
Now, a
 diagnosis is a set of endogenous tuples that, by becoming abnormal, restore consistency.
 \begin{definition}\label{def:diag}
 (a) A {\em diagnosis} for a diagnosis problem $\mc{M}$ is a $\Delta \subseteq D^n$, such that
\ $\nit{SD} \cup \{\nit{ab}_P(\bar{c})~|~P(\bar{c}) \in \Delta\} \cup \{\neg \nit{ab}_P(\bar{c})~|~P(\bar{c}) \in D \smallsetminus \Delta\} \cup \{\mc{Q}\}$ \ becomes consistent.
\ (b)
$\mc{D}(\mc{M},t)$ denotes the set of subset-minimal diagnoses for $\mc{M}$ that contain a tuple $t \in D^n$. \ (c) $\mc{MCD}(\mc{M},t)$ denotes the set of  diagnoses of
$\mc{M}$ that contain a tuple $t \in D^n$ and have the minimum cardinality (among those diagnoses that contain $t$). \boxtheorem
\end{definition}
Clearly,
$\mc{MCD}(\mc{M},t) \subseteq \mc{D}(\mc{M},t)$. The following proposition specifies the relationship between minimal diagnoses for $\mc{M}$
and actual causes for $\mc{Q}$.

\begin{proposition}\label{pro:ac&diag}
Consider  $D= D^n \cup D^x$, a BCQ $\mc{Q}$, and the diagnosis problem $\mc{M}$ associated to $\mc{Q}$. Tuple $t \in D^n$ is an actual cause for $\mc{Q}$
iff $\mc{D}(\mc{M},t) \not = \emptyset$. \boxtheorem
\end{proposition}
\ignore{To establish a connection between cardinality of diagnoses for $\mc{M}$ and responsibility for actual causes of $\mc{Q}$, let $\mc{MCD}(\mc{M},t)$ denote the set of  diagnoses of
$\mc{M}$ that contain a tuple $t \in D^n$ and have the minimum cardinality (among those diagnoses that contain $t$). }
The next proposition tells us that the responsibility of an actual cause $t$ is determined by the cardinality of the diagnoses in $\mc{MCD}(\mc{M},t)$.

% i.e., $\mc{MCD}(\mc{M},t)=$ $  \{ \Delta| \Delta$ is a
%minimal diagnoses for $\mc{M}$ such that there is no minimal diagnosis such as $\Delta'$ for $\mc{M}$  for which  $t \in \Delta'$ and $|\Delta'| <|\Delta|$ and $t \in \Delta'  \}$

\begin{proposition}\label{pro:r&diag}
Consider  $D= D^n \cup D^x$, a BCQ $\mc{Q}$, the diagnosis problem $\mc{M}$ associated to $\mc{Q}$, and a tuple $t \in D^n$.
%Given a database $D= D^x \cup D^n$, a BCQ $\mc{Q}$ and a tuple $t\in D$,
\vspace{-1mm}
\begin{itemize}
\item[(a)] $\rho(t)=0$ iff $\mc{MCD}(\mc{M},t) = \emptyset$.
\item [(b)] Otherwise, $\rho(t)=\frac{1}{|s|}$, where $s \in \mc{MCD}(\mc{M},t)$. \boxtheorem
\end{itemize}
\end{proposition}
\vspace{-3mm}
\begin{example}\label{ex:mbdaex6}   (ex. \ref{ex:mbdaex5} cont.) The diagnosis problem $\mc{M}$ has two diagnosis namely,
$\Delta_1=\{ S(a_3)\}$ and $\Delta_4=\{ S(a_4)\}$.

Here, $\mc{D}(\mc{M},S(a_3))= \mc{MCD}(\mc{M}, S(a_3))=\{ \{ S(a_3)\}\}$ and $\mc{D}(\mc{M}, S(a_4))=\mc{MCD}(\mc{M}, S(a_4))=\{\{ $ $S(a_4)\}\}$. Therefore, according to Proposition
\ref{pro:ac&diag} and \ref{pro:r&diag}, both $S(a_3)$ and $S(a_4)$ are actual cases for $\mc{Q}$, with responsibility 1. \boxtheorem
\end{example}
Notice that the consistency-based approach to causality provided in this section can be considered as a technique for computing repairs for
inconsistent databases wrt.\ denial constraints (it is a corollary of \ref{pro:sr&cp} and \ref{pro:r&diag}). It is worth mentioning that this approach has been implicitly used before
in databases repairing in \cite{Arenas03}, where
the authors introduce {\em conflict graphs} to characterize S-repairs for inconsistent databases wrt.\ FDs. We will use this connection in our future work to provide some complexity
results in the context of causality.

\ignore{
\vspace{-3mm} \subsection{ Consistency-Based Reasoning and Database Repairing} \label{sec:MBDtoRep}

The consistency-based framework that developed in this section can be regarded as a framework to compute repairs for inconsistent databases wrt.\ denial constraint. Suppose
we are given an inconsistent database $D$ wrt.\ a denial constraint $\psi$. Consider the set consistency-based diagnosis
system  $\mathcal{M}=(\nit{Th(D)},D^n,\psi^{Q^{ex}})$. The following proposition shows that diagnoses of $M$ characterize both S.re[pairs and C.repairs of the database.

\noindent

It is worth mentioning that consistency-based reasoning have been used before implicity in the context of databases repairing in \cite{Arenas03} when
the authors introduced {\em conflict graphs} to characterize S.repairs for inconsistent databases wrt.\ FDs.

\noindent  Given an inconsistent database $D$ wrt.\ a FD $\psi$, the conflict graph $\mathcal{G}_{D,\psi}$ have as vertices
the database tuples; and edges connect two tuples that simultaneously violate
a FD. There is a one-to-one correspondence between S-repairs of the database
and the set-theoretically maximal independent sets in the conflict graph \cite{Arenas03}. Similarly, there is a one-to-one correspondence between C-repairs and maximum
independent sets in the same graph. Conflict graphs for databases wrt general denial constraints become conflict
hypergraphs \cite{Chomicki05}.

\noindent We show that there is a direct relationship between conflict graphs hypergraph for inconsistent databases wrt denial
 constraint and conflict sets in our consistency-based diagnosis approach to compute repairs.

Let $\mathcal{C}(M)$ denotes the collection of conflict sets for $M$. According to
the Theorem \ref{the:diag} any hitting set over $\mathcal{C}(M)$ is a diagnosis for $M$. The hitting set problem over a collection can be viewed as the {\em vertex cover problem}
on {\em  hypergraphs}.  Interpret the elements of $S$ as vertices and
interpret the subsets of $S$ as hyperedges. It is not difficult to show that the corresponding hypergraph of the collection $\mathcal{C}(M)$ is isomorphic to  $\mathcal{G}_{D,\psi}$.
In the former any minimal vertex cover of  $H_M$ must removed from the instance to
obtain a S.repair in the latter any maximal independent set forms a repair. We know minimal
vertex cover and maximal independent set are dual notions, so basically both approach are the same.
}

\section{Discussion}\label{sec:disc}

Here we discuss some directions of possible or ongoing research.

\vspace{-3mm}
\paragraph{Open queries.} \ We have limited our discussion to boolean queries. It is possible to extend our work
to consider conjunctive queries with free variables, e.g. $\mc{Q}(x)\!: \exists yz(R(x,y) \wedge S(y,z))$. In this case,
a query answer would be of the form $\langle a\rangle$, for $a$ a constant, and causes would be found for such an answer.
In this case, the associated denial constraint would be of the form $\kappa^{\langle a\rangle}\!: \ \leftarrow R(a,y), S(y,z)$, and
the rest would be basically as above.

\vspace{-4mm}
\paragraph{Algorithms and complexity.} \ Given the connection between causes and different kinds of repairs, we might take advantage
for causality of algorithms and complexity results obtained for database repairs. This is matter of our ongoing research. In this work, apart
from providing a naive algorithm for computing repairs from causes, we have not gone into detailed algorithm or complexity issues. The results
we already have in this direction will be left for an extended version of this work.

\vspace{-4mm}
\paragraph{Endogenous repairs.} \ The partition of a database into endogenous and exogenous tuples has been exploited in the
context of causality. However, this kind of partition is also of interest in the context of repairs. Considering that we should
have more control on endogenous tuples than on exogenous ones, which may come from external sources, it makes sense to consider
{\em endogenous repairs} that are obtained by updates (of any kind) on endogenous tuples. For example, in the case of violation of denial constraints,
   endogenous repairs would be obtained -if possible- by deleting endogenous tuples only. If there are no repairs based on endogenous tuples
   only, a preference condition could be imposed on repairs \cite{ihab12,chomicki12}, privileging those that change exogenous the least. (Of course,
   it could also be the other way around, that is we may feel more inclined to change exogenous tuples than our endogenous ones.)

   As a further extension, it could be possible to assume that combinations of (only) exogenous tuples never violate the ICs, something that could be checked
   at upload time. In this sense, there would be a part of the database that is considered to be consistent, while the other is subject to possible repairs.
   A situation like this has been considered, for other purposes and in a different form, in \cite{greco14}.

  Actually, going a bit further, we could even consider the relations  in the database with an extra, binary  attribute, $N$, that is used to annotate if a tuple is
  endogenous or exogenous (it could be both), e.g. a tuple like $R(a,b, \nit{yes})$. ICs could be annotated too, e.g. the ``exogenous" version of DC $\kappa$, could be
   $\kappa^E\!: \ \leftarrow P(x, y,\nit{yes}),R(y, z,\nit{yes})$, and could be assumed to be satisfied.

\vspace{-4mm}
   \paragraph{ASP specification of causes.} \ Above we have presented a connection between causes and repairs. S-repairs can be specified by means of
   answer set programs (ASPs) \cite{tplp03,barcelo02,barcelo03}, and C-repairs too, with the use of weak program constraints \cite{tplp03}. This should allow for the
   introduction of ASPs in the context of causality, for specification and reasoning.
    There are also ASP-based specifications of diagnosis \cite{eiter99} that could be brought into a more complete picture.

  \ignore{  \paragraph{Causes for IC violations.} \ We may consider a set $\Sigma$ of ICs $\psi$ that have violation views $V^\psi$ that become boolean conjunctive queries, e.g. denial constraints.
    Each of such views has the form $V^\psi \!: \ \exists \bar{x} (A_1(\bar{x}_1) \wedge \cdots \wedge A_n(\bar{x}_n))$. When the instance $D$ is inconsistent wrt.\ $\Sigma$, some of these
    views (queries) get the answer \nit{yes} (they become true), and for each of them there is a set $\mc{C}(D,D^n,V^\psi)$ whose elements are of the form $\langle t, \{C_1(t), \ldots, C_m(t)\}\rangle$, where $t$
    is a tuple that is an  actual cause for $V^\psi$, together with their contingency sets $C_i(t)$,
    possibly minimal in some sense. The question is how can we obtain repairs of $D$ wrt.\ $\Sigma$ from the sets $\mc{C}(D,D^n,V^\psi)$? How does consistent query answering appear here and what does it have to say about
    causality? }

\vspace{-4mm}
    \paragraph{Causes and functional dependencies.} \ Functional dependencies (FDs), that can be considered as denial constraints, have violation views that are conjunctive, but
    contain inequalities.   They are still monotonic views though. Much has been done in the area of repairs and consistent query answering \cite{2011Bertossi}. On the other side, in causality only conjunctive queries without built-ins have been considered \cite{Meliou2010a}.
    It is possible that causality can be extended to conjunctive queries with built-ins through the repair connection; and also to non-conjunctive queries via repairs wrt.\ more complex
    integrity constraints.

    \vspace{-3mm}
    \paragraph{View updates.} \ Another  venue to explore for fruitful connections relates to the {\em view update problem}, which is about updating a database through views.
This old and important problem in databases has also been treated from the point of view of abductive reasoning
\cite{Kakas90,Console95}.\footnote{Abduction has also been explicitly applied to database repairs \cite{arieli}.} User knowledge imposed through view updates creates or reflects {\em uncertainty} about the base data, because alternative base instances may give an account
of the intended view updates.

The view update problem, specially in its particular form of of {\em deletion propagation}, has been recently related in \cite{benny12a,benny12b} to causality as introduced in
\cite{Meliou2010a}.\footnote{Notice only tuple deletions are used with violation views and repairs associated to denial constraints.}

Database repairs are also related to the view update problem.
Actually, {\em answer set programs} (ASP) for database repairs \cite{barcelo03} implicity repair the database by updating intentional,
annotated predicates.

Even more, in \cite{lechen}, in order to protect sensitive information, databases are explicitly and virtually ``repaired" through secrecy views that specify the
information that has to be kept secret. In order to protect
information, a user is allowed to interact only with the virtually repaired versions of the original database that result from making those views empty or
contain only null values. Repairs are specified and computed using ASP, and in \cite{lechen} an explicit connection to prioritized attribute-based repairs \cite{2011Bertossi} is made.

\section{Conclusions} \label{sec:abdcomx}

In this work, we have uncovered  the relationships between causality in databases, database repairs, and consistency-based reasoning, as three forms of non-monotonic reasoning. Establishing the connection between these problems allows us to apply results and techniques developed for
each of them to the others. This should  be particularly beneficial for causality in databases, where still a  limited number of results and techniques have been obtained or developed.
This becomes matter of our ongoing and future research.

 Our work suggests that diagnostic reasoning, as a form of non-monotonic reasoning, can provide a  solid theoretical foundation  for query answer explanation and provenance. The need
  for such foundation and the possibility of using non-monotonic logic for this purpose
are mentioned in \cite{Cheney09b,Cheney11}. %Providing a semantic for provenance i   n database based on diagnostic reasoning is our future work.

\vspace{1mm}
\noindent {\bf Acknowledgments:} \ Research funded by NSERC Discovery, and
the NSERC Strategic Network on Business Intelligence (BIN). L.
Bertossi is a Faculty Fellow of IBM CAS. Conversations on causality in databases with Alexandra Meliou during Leo Bertossi's visit to U. of Washington in 2011 are much appreciated.
He is also grateful to Dan Suciu and  Wolfgang Gatterbauer for their hospitality. Leo Bertossi is also grateful to Benny Kimelfeld for stimulating conversations at LogicBlox, and pointing out to \cite{benny12a,benny12b}.

\ignore{
\comlb{I still do not like your bibliography. First all all it has to be homogeneous, in particular, all titles
in lower case or upper case (see my papers). Journal names are emphasized. Book titles too. Yo do not add "J." in front
of a journal name that does not have it. For example, "J. Th. Comp. Sci" does not exist. Delete it from  where it should not be.
You do not add a colon after the authors names.
Even less if you do only sometimes. The same author cannot have two different first name initials.  You do not add "Conf" after a conference
unless it is part of the title. In particular, it is not "Sigmod Conf". For books add the publisher. For journal publications, if you have the
volume (say 4) and the issue (say 3) (which you should try to provide), and the pages, you add after the year: "4(3):345-360". The "In" almost everywhere has to be eliminated.
Most importantly, look at my papers.}

\comlb{More specifically, now, these are citations to consider:

- Halpern, Y.~J., Pearl, .J. Causes and Explanations: A Structural-Model Approach: Part 1. British J. Philosophy of Science, 2005, 56:843-887.

- Why do you have so many entries for Cheney? Have you read them all? Do they differ much? You need them?

- What about Tannen? I told you to find something by him, not to eliminate him!}
\combabak{ \em I think we need all the papers by cheney, \cite{Cheney09} is an important introduction to provenance, \cite{Cheney09b, Cheney11} are critically important for our future work. (to use non-monotonic logic to provide a semantic for provenance)
I do read all of these papers btw.

I deleted Tannen because I could not find a paper by him which provide a general picture of provenance. Anyways, you might want to cite \cite{Tannen10} which I happen to read it.}

}

\ignore{

\newpage
\section*{Proofs of Results}\label{ap:proofs}

\defproof{Proposition \ref{pro:nonmon}}{(a) \ Assume an arbitrary element $t' \in  \mc{CS}(D^n,D^x,\mc{Q})$. Since $t'$ is an actual cause for $\mc{Q}$ then there exist a set $\Gamma \subseteq D^n$
such that: $D^x \cup D^n \smallsetminus  \Gamma \models \mc{Q}$ and  $D^x \cup D^n \smallsetminus  (\Gamma \cup \{t\}) \not \models \mc{Q}$. Now consider $U(D)= (D^n)^\prime \cup D^x$.
Let $\Gamma'=\Gamma \cup \{t\}$ then, $D^x \cup (D^n)^\prime \smallsetminus  \Gamma' \models \mc{Q}$ and
$D^x \cup (D^n)^\prime \smallsetminus  (\Gamma \cup \{t\}) \not \models \mc{Q}$ (because, $(D^n)^\prime \smallsetminus  \Gamma' =D^n \smallsetminus  \Gamma$). So, we obtain that $t'$ is counterfactual cause for $\mc{Q}$. Consequently, $\mc{CS}(D^n,D^x,\mc{Q})$ $\subseteq \mc{CS}((D^n)^\prime,D^x,\mc{Q})$.

\noindent (b) \
 We show that inserting an exogenous tuples do not add any new actual cause for a query answer. Assume the contradiction i.e., there exists a $t \in D^n$ such that $t \in \mc{CS}(D^n, (D^x)^\prime, \mc{Q})$ but
$t \not \in \mc{CS}(D^n, D^x, \mc{Q})$. The former statement implies that: there exists a set $\Gamma \subseteq D^n$ such that $ D^n \cup (D^x)^\prime \smallsetminus \Gamma \models \mc{Q}$
and  $ D^n \cup (D^x)^\prime \smallsetminus (\Gamma \cup \{ t \}) \not \models \mc{Q}$. The latter, implies that for all $s \subseteq D^n$ (including $\Gamma$ and $\emptyset$) we have
$(a): D^n \cup D^x \smallsetminus s \not \models \mc{Q}$ or $(b):D^n \cup D^x \smallsetminus (s \cup \{ t \}) \models  \mc{Q}$. We show that both (a) and (b) lead to contradiction.
When $s=\emptyset$, (a) is equal to $D^n \cup D^x \not \models \mc{Q}$ i.e., $ D \not \models \mc{Q}$ (contradiction!). When $s=\Gamma$, (b)
is equal to $D^n \cup D^x \smallsetminus (\Gamma \cup \{ t \}) \models  \mc{Q}$ which contradict the fact that
 $ D^n \cup (D^x)^\prime \smallsetminus (\Gamma \cup \{ t \}) \not \models \mc{Q}$, since $\mc{Q}$ is a
 monotone query. Therefore, $ \mc{CS}(D^n, (D^x)^\prime, \mc{Q}) \subseteq \mc{CS}(D^n,D^x,\mc{Q}) $.  }

\section*{The Algorithm}

 \begin{algorithm}
\caption{ Computing repairs from actual causes }

 \label{alg:srep}
\begin{algorithmic}[1]

\Function{SubsetRepairs}{$D$, $\Sigma$,  $\mc{C}(D,V^{\psi})$}
\State $S \leftarrow \{ t \; | \; \exists \psi \in \Sigma, t \in \mc{CS}(D, V^\psi)\}$
\State $C \leftarrow \{ \langle \mc{CS}(D, V^\psi) \rangle \; | \; \exists \psi \in \Sigma, \mc{CS}(D, V^\psi)  \not = \emptyset  \}$
\State $ \mc{H} \leftarrow$ \Call{MinHittingSets}{$S$,$C$}
\State $\sigma=\{ \psi \; | \psi  \in \Sigma, D \models V^\psi    \}$
\State $\nit{Reps} \leftarrow \emptyset$
\ForAll{$h \in \mc{H} $}
\ForAll{$c \in$ \Call{Cont}{$h$, $\sigma$} }
\State  Add $ D \smallsetminus c$ to  the collection $\nit{Reps}$
\EndFor
\EndFor

\Return   $\nit{Reps}$
\EndFunction

\Function{Cont}{$h$, $\sigma$}
\ForAll{$\psi_i \in \sigma$}
 \State  $S_i \leftarrow \emptyset $
 \ForAll{ $t \in h$}
  \If{ $t \in \mc{CS}(D, V^{\psi_i}) $ }

   \State Add $  \bigcup_{j=1 \cdots m} \langle  t, C_j(t)\rangle$ to the set $S_i$ where, \\
   \hspace{1.8cm} $\langle t, \{C_1(t), \ldots, C_m(t)\} \rangle \in \mc{C}(D,V^{\psi_i})$
  \EndIf
 \EndFor
\EndFor

\Return  $\{ \langle \bigcup_{i=1 \cdots n} c_i \rangle |  c_i \in S_i \}$

\EndFunction

\end{algorithmic}

\end{algorithm}

}


\begin{thebibliography}{10}
\bibliographystyle{aaai}

\bibitem[Arenas, Bertossi, and Chomicki 1999]{pods99}
Arenas, M., Bertossi, L. and Chomicki, J. \newblock Consistent Query Answers in Inconsistent Databases. \newblock {\em Proc. ACM PODS}, 1999, pp. 68-79.


\bibitem[Arenas, Bertossi, and Chomicki 2003]{tplp03}
Arenas,~M., Bertossi,~L., Chomicki,~J. \newblock Answer Sets for Consistent Query Answers. {\em Theory and Practice of Logic Programming}, 2003, 3(4\&5):393-424.

\bibitem[Arenas et al. 2003]{Arenas03}
Arenas, M., Bertossi, L., Chomicki, J., He, X., Raghavan, V. and Spinrad, J.
\newblock Scalar Aggregation in Inconsistent Databases.
\newblock {\em Theoretical Computer Science}, 2003, 296:405-434.

\bibitem[Arieli et al. 2004]{arieli}
Arieli,~O., Denecker,~M., Van Nuffelen,~B. and Bruynooghe,~M. \newblock Coherent Integration
of Databases by Abductive Logic Programming. {\em J. Artif. Intell. Res.}, 2004, 21:245-286.

\bibitem[Barcelo, and Bertossi 2002]{barcelo02}
Barcelo, P. and Bertossi, L. \newblock Repairing Databases with Annotated Predicate Logic. {\em Proc. NMR}, 2002.

\bibitem[Barcelo, Bertossi, and Bravo 2003]{barcelo03}
Barcelo,~P., Bertossi,~L. and Bravo,~L. \newblock Characterizing and Computing Semantically Correct Answers from Databases with Annotated Logic and Answer Sets.
In {\em Semantics of Databases}, Springer LNCS 2582, 2003, pp. 1-27.


\bibitem[Bertossi, and Li 2013]{lechen}
Bertossi,~L. and Li,~L. \newblock Achieving Data Privacy through Secrecy Views and Null-Based Virtual Updates. {\em IEEE Transaction on Knowledge and Data Engineering}, 2013, 25(5):987-1000.

\bibitem[Bertossi 2011]{2011Bertossi}
Bertossi, L.
\newblock {\em Database Repairing and Consistent Query Answering}.
\newblock Morgan \& Claypool, Synthesis Lectures on Data Management, 2011.


\bibitem[Bertossi 2006]{Bertossi06}
Bertossi, L.
\newblock Consistent Query Answering in Databases.
\newblock {\em ACM SIGMOD Record},  2006, 35(2):68-76.

\bibitem[Borgida, Calvanese, and Rodriguez-Muro 2008]{borgida}
Borgida, A., Calvanese, D. and Rodriguez-Muro, M. Explanation in DL-Lite.
{\em Proc. DL WS}, CEUR-WS 353, 2008.


\bibitem[Buneman, Khanna, and Tan 2001]{BunemanKT01}
Buneman, P., Khanna, S. and Tan, W.~C.
\newblock Why and Where: A Characterization of Data Provenance.
\newblock {\em Proc. ICDT}, 2001, pp. 316--330.

\bibitem[Buneman, and Tan 2007]{BunemanT07}
Buneman, P. and Tan, W.~C.
\newblock Provenance in Databases.
\newblock {\em Proc. ACM SIGMOD}, 2007,  pp. 1171--1173.


\bibitem[Chapman, and Jagadish 2009]{ChapmanJ09}
 Chapman, A., and  Jagadish, H.~V.
\newblock Why Not?
\newblock {\em Proc. ACM SIGMOD}, 2009, pp.523--534.

\bibitem[Cheney, Chiticariu,  and  Tan 2009]{Cheney09}
Cheney, J., Chiticariu, L. and  Tan, W.~C.
\newblock Provenance in Databases: Why, How, And Where.
\newblock {\em Foundations and Trends in Databases}, 2009, 1(4): 379-474.


\bibitem[Cheney et al. 2009]{Cheney09b}
Cheney, J., Chong, S., Foster, N., Seltzer, M.~I. and Vansummeren, S.
\newblock Provenance: A Future History.
\newblock {\em OOPSLA Companion (Onward!)}, 2009, pp. 957--964.

\bibitem[Cheney 2011]{Cheney11}
Cheney, J.
\newblock Is Provenance Logical?
\newblock {\em Proc. LID}, 2011, pp. 2--6.

\bibitem[Chomicki, and Marcinkowski 2005]{Chomicki05}
 Chomicki, J. and Marcinkowski, J.
\newblock Minimal-Change Integrity Maintenance Using
Tuple Deletions.
\newblock {\em Information and Computation}, 2005, 197(1-2):90-121.

\bibitem[Chockler, and Halpern 2004]{cs-AI-0312038}
 Chockler, H.  and Halpern,  J.~Y.
\newblock Responsibility and Blame: A Structural-Model Approach.
\newblock {\em J. Artif. Intell. Res.}, 2004, 22:93-115.

\bibitem[Console, Sapino, and Theseider-Dupre 1995]{Console95}
 Console, L., Sapino M.~L., Theseider-Dupre,~D.
\newblock   The Role of
Abduction in Database View Updating.
\newblock {\em  J. Intell. Inf. Syst.}, 1995, 4(3): 261-280.

\bibitem[Cui, Widom, and  Wiener 2000]{CuiWW00}
Cui, Y., Widom,  J.  and  Wiener, J.~L.
\newblock Tracing The Lineage of View Data in a Warehousing Environment.
\newblock {\em ACM Trans. Database Syst.}, 2000, 25(2):179-227.

\bibitem[Eiter et al. 1999]{eiter99}
Eiter,~Th., Faber,~W., Leone,~N. and Pfeifer,~G. \newblock The Diagnosis Frontend of the DLV System. {\em AI Commun.}, 1999, 12(1-2):99-111.

\bibitem[Gertz 1997]{Gertz97}
 Gertz, M.
\newblock Diagnosis and Repair of Constraint Violations in Database Systems.
\newblock  PhD Thesis, Universit\"at Hannover, 1996.


\bibitem[Greco, Pijcke, and Wijsen 2014]{greco14}
Greco,~S., Pijcke,~F. and Wijsen,~J. \newblock Certain Query Answering in Partially Consistent Databases. {\em PVLDB}, 2014, 7(5):353-364.

\bibitem[Halpern, and Pearl 2001]{Halpern01}
Halpern, Y.~J., and Pearl, J.
\newblock Causes and Explanations: A Structural-Model Approach: Part 1
\newblock {\em Proc. UAI}, 2001, pp. 194-202.

\bibitem[Halpern, and Pearl 2005]{Halpern05}
Halpern, Y.~J., and Pearl, J.
\newblock Causes and Explanations: A Structural-Model Approach: Part 1.
\newblock {\em British J. Philosophy of Science}, 2005,
56:843-887.

\bibitem[Huang et al. 2008]{HuangCDN08}
 Huang, J.,  Chen, T.,  Doan, A. and  Naughton, J.~F.
\newblock On The Provenance of Non-Answers to Queries over Extracted Data.
\newblock {\em PVLDB}, 2008, 1(1):736--747.

\bibitem[Kakas, and   Mancarella 1990]{Kakas90}
Kakas A.~C. and   Mancarella,~P.
\newblock  Database Updates through Abduction.
\newblock {\em Proc. VLDB}, 1990, pp. 650-661.

\bibitem[Karvounarakis, and Green 2012]{Karvounarakis02}
Karvounarakis, G. and Green, T.~J.
\newblock Semiring-Annotated Data: Queries and Provenance?
\newblock {\em SIGMOD Record}, 2012, 41(3):5-14.

\bibitem[Karvounarakis, Ives, and Tannen 2010]{Tannen10}
Karvounarakis, G.   Ives, Z.~G. and Tannen, V.
\newblock  Querying Data Provenance.
\newblock {\em Proc. ACM SIGMOD}, 2010, pp. 951--962.

\bibitem[Kimelfeld 2012]{benny12a}
Kimelfeld, B. A Dichotomy in the Complexity of Deletion Propagation with Func-
tional Dependencies.
Proc. ACM PODS, 2012.

\bibitem[Kimelfeld, Vondrak, and Williams 2012]{benny12b}
Kimelfeld, B., Vondrak, J. and Williams, R. Maximizing Conjunctive Views in
Deletion Propagation.
ACM Trans. Database Syst., 2012, 37(4):24.

\bibitem[Lopatenko, and Bertossi 2007]{icdt07}
Lopatenko, A. and Bertossi, L. \newblock Complexity of Consistent Query Answering in Databases under Cardinality-Based and Incremental Repair Semantics. \newblock {\em Proc. ICDT}, 2007, Springer LNCS 4353, pp. 179-193.

\bibitem[Meliou et al. 2010a]{Meliou2010a}
Meliou, A.,  Gatterbauer, W.
 Moore, K.~F. and  Suciu, D.
\newblock The Complexity of Causality and Responsibility for Query Answers and
  Non-Answers.
\newblock {\em Proc. VLDB}, 2010, pp. 34-41.

\bibitem[Meliou et al. 2010b]{Meliou2010b}
Meliou, A.,  Gatterbauer. W., Halpern,  J.~Y.,  Koch, C.,
 Moore K.~F. and  Suciu, D.
\newblock Causality in Databases.
\newblock {\em IEEE Data Eng. Bull}, 2010, 33(3):59-67.

\bibitem[Reiter 1987]{Reiter87}
 Reiter, R.
\newblock A Theory of Diagnosis from First Principles.
\newblock {\em Artificial Intelligence}, 1987, 32(1):57-95.

\bibitem[Reiter 1982]{Reiter82}
Reiter, R.
\newblock Towards a Logical Reconstruction of Relational Database Theory. \newblock
 In {\em On Conceptual Modelling}, \ignore{M. Brodie, J. Mylopoulos and J.W. Schmidt (eds.),} Springer, 1984,
pp. 191-233.


\bibitem[Staworko, Chomicki, and Marcinkowski 2012]{chomicki12}
Staworko,~S., Chomicki,~J. and Marcinkowski,~J. \newblock Prioritized Repairing and Consistent Query Answering in Relational Databases. \newblock {\em Ann. Math. Artif. Intell.}, 2012, 64(2-3):209-246.

\bibitem[Struss 2008]{struss}
Struss, P. \newblock Model-based Problem Solving. \newblock In {\em Handbook of Knowledge Representation},
chapter 10. Elsevier, 2008.

\bibitem[Tannen 2013]{tannen}
Tannen, V. \newblock Provenance Propagation in Complex Queries. \newblock In {\em Buneman Festschrift}, 2013, Springer LNCS 8000, pp. 483"1¤73.

\bibitem[Yakout et al. 2011]{ihab12}
Yakout,~M., Elmagarmid,~A., Neville,~J., Ouzzani,~M. and Ilyas,~I. \newblock Guided Data Repair. {\em PVLDB}, 2011, 4(5):279-289.

\end{thebibliography}
\end{document}